\documentclass{article}

\usepackage{amssymb,amsfonts,amsmath}
\usepackage{graphicx}
\usepackage{subfig}
\usepackage{bm}
\usepackage[round]{natbib}
\usepackage{url, color}
\usepackage{authblk}

\begin{document}

\title{Bayesian identification of bacterial strains from sequencing data}

\author[1]{Aravind Sankar}
\author[1]{Brandon Malone}
\author[2]{Sion Bayliss}
\author[3]{Ben Pascoe}
\author[3]{Guillaume M\'{e}ric}
\author[3]{Matthew D. Hitchings}
\author[3]{Samuel K. Sheppard}
\author[2]{Edward J. Feil}
\author[4]{Jukka Corander}
\author[1]{Antti Honkela}

\affil[1]{Helsinki Institute for Information
    Technology HIIT, Department of Computer Science, University of
    Helsinki, Helsinki, Finland}
\affil[2]{Department of Biology and Biochemistry, University of Bath, United Kingdom}
\affil[3]{Institute of Life Sciences, College of Medicine, Swansea University, United Kingdom}
\affil[4]{Helsinki Institute for Information Technology HIIT,
    Department of Mathematics and Statistics, University of
    Helsinki, Helsinki, Finland}

\date{}

\maketitle

\begin{abstract}
 Rapidly assaying the diversity of a bacterial species present in a sample obtained from a hospital patient or an evironmental source has become possible after recent technological advances in DNA sequencing. For several applications it is important to accurately identify the presence and estimate relative abundances
  of the target organisms from short sequence reads obtained from a
  sample. This task is particularly challenging when the set of
  interest includes very closely related organisms, such as different
  strains of pathogenic bacteria, which can vary considerably in terms
  of virulence, resistance and spread. Using advanced Bayesian
  statistical modelling and computation techniques we introduce a
  novel pipeline for bacterial identification that is shown to
  outperform the currently leading pipeline for this purpose. Our
  approach enables fast and accurate sequence-based identification of
  bacterial strains while using only modest computational
  resources. Hence it provides a useful tool for a wide spectrum of
  applications, including rapid clinical diagnostics to distinguish
  among closely related strains causing nosocomial infections. The
  software implementation is available at \url{https://github.com/PROBIC/BIB}.
\end{abstract}

\section{Introduction}

Different strains of pathogenic bacteria are known to often vary in
terms of virulence, resistance and geographical spread~\citep{Meric2015}. Rapid and inexpensive sequence-based identification of the strain(s)
colonising a patient would be highly desirable. Previous research shows that patients can often
host several strains of specific \emph{Staphylococcus}
species~\citep{Ueta2007}. The current approach is to isolate single colonies assuming that the sample is homogeneous. Here we consider an approach which allows a robust test
of this assumption, and additionally, if there is diversity, an
efficient means to compare similarities between whole host populations
present in different individuals. Since single random colonies might
be misleading, it is beneficial to allow for a more flexible approach
where pooled colony data can be directly utlized.

With the growing tendency to routinely sequence
samples from infected patients in the hospital environment, the
identification would be additionally advantageous for pathogen
surveillance and monitoring
purposes without necessitating the use of extensive computational
resources for \textit{de novo} genome assembly. Moreover, samples with
mixed presence of several strains are problematic for assembly-based
analyses, which calls for alternative approaches.

The identification of bacteria from sequencing data has been widely
considered in metagenomic community profiling~\citep{Segata2013,Franzosa2015}.
As our primary identification and estimation focus is at a much higher level of resolution than in typical metagenomics studies, whole genome or whole
metagenome shotgun sequencing data is by definition a necessity for a successful implementation of a platform for this purpose. Typical metagenomic approaches for such data are based on defining a set of markers for
each clade of interest~\citep{Segata2012,Sunagawa2013}. However, these methods
are typically not sensitive enough to identify the pathogens
responsible for infections in sufficient detail. \citet{Eyre2013}
present a method for detecting mixed infections but the method assumes
there are at most two strains in each sample, which may not hold, in
particular if a sample has become contaminated at any phase of the
preparation and sequencing process. A Bayesian
statistical method capable of using all the sequencing data was
recently introduced~\citep{Francis2013,Hong2014}, but also its
practical performance may not be appropriate as suggested by our experiments.

The computational problem in bacterial strain identification is
analogous to the widely studied transcript isoform expression
estimation in RNA-sequencing (RNA-seq) data analysis, which both
aim at identifying and quantifying the abundance of several closely
related sequences from short read data. In both cases
a significant fraction of reads will align perfectly to multiple
sequences of interest. Several probabilistic models have been
proposed for solving this problem~\citep{Xing2006,Jiang2009}.
Based on its success in recent
assessments of methods for this problem~\citep{SEQC2014,Kanitz2015}, we
use the BitSeq~\citep{Glaus2012,Hensman2015} method to obtain a fast
and accurate solution to this problem in our Bayesian Identification
of Bacteria (BIB) pipeline for bacterial strain
identification from unassembled sequence reads.

In this paper we focus on
\textit{Staphylococcus aureus} and \textit{S. epidermidis}, which
represent two of the most widespread causes of nosocomial infections
and impose considerable burden on the public health system
worldwide~\citep{Harris2010,Meric2015}. Using a diverse collection of
strains from these two species as a model system, we demonstrate that
clinically relevant, fast and highly accurate identification of the
strains colonising a patient is possible in less than 10 minutes on
a standard single CPU desktop computer.
Our BIB pipeline improves significantly upon the
state-of-the-art approach for sequence based identification of
bacteria.

\section{Materials and methods}

Our pipeline is built by a combination of the following two central ideas:
\begin{enumerate}
\item defining core genomes of the target set of strains by excluding
  more variable regions to strengthen the analysis, and
\item using a fast fully probabilistic method to estimate the relative frequencies of
  the target strains in a sample.
\end{enumerate}

These ideas translate to two analysis steps:
\begin{enumerate}
\item[Step 1] Cluster the strains, perform multiple sequence alignment
  to find the strain-specific common core genome and construct an
  index for read alignment.
\item[Step 2] Align the reads to the reference core genomes allowing
  multiple matches and use a probabilistic method to estimate the
  strain abundances using the alignments.
\end{enumerate}
Step 1 only needs to be done once for each collection of reference
sequences while Step 2 needs to be performed for every sample.
The two steps will be detailed further below, followed by description of the
synthetic data generation process and characterisation of the real
data which are used for empirical evaluation and comparison against the
leading alternative identification method.

\subsection{Step 1: Reference strain selection and core genome extraction}

We demonstrate our pipeline on a collection of 30 \emph{S. aureus} and
3 \emph{S. epiderimdis} strains whose phylogenetic tree is illustrated
in Fig.~\ref{fig:tree}.  The tree was constructed using UPGMA method
with p-distance in the MEGA6 software \citep{Tamura2013}. The tree
displays a natural partition with 13 \emph{S.\ aureus} strain
clusters, each of which corresponds to an already established clonal
complex \citep{Feil2004}, while each \emph{S.\ epidermidis} strain
forms a cluster of its own, representing the three previously
identified main complexes within the species \citep{Meric2015}.  The
strains selected to represent each cluster are in bold in
Fig.~\ref{fig:tree}.

Microbial genomes are often highly dynamic and susceptible to horizontal gene
transfer and translocation of genomic regions~\citep{Gogarten2002,Lawrence2002}.
As a consequence, mobile elements may
confuse genome-based identification of strains. In order to avoid issues
with misalignment of reads and incorrect abundance estimates, we
discard the non-core parts of the reference genomes and use only core
alignment, i.e.\ part of the genome shared by all strains of a
species, as a basis for the analysis.

A multiple sequence alignment for the 16 cluster prototype bacterial strains shown in bold in Fig.~\ref{fig:tree} was obtained using progressiveMauve~\citep{Darling2010}. The accessory
genome regions were detected and discarded using the standard settings, resulting in an ungapped core
alignment which was used to represent the genomic variation in the target set of strains.
These ungapped sequences are used to construct an index for read alignment.

\subsection{Step 2: Strain abundance estimation}

The gapless core genomes extracted as described above were considered as the reference
sequences in the BitSeq~\citep{Glaus2012,Hensman2015} method to estimate
the relative proportion of each strain in our reference collection in a sample. We used Bowtie 2~\citep{Langmead2012}
to align the reads to the reference sequences allowing for multiple
matches. We then used \texttt{estimateVBExpression} from BitSeq to
estimate the relative proportions of each of the strains in the sequenced samples. Our full method pipeline is referred to as Bayesian Identification of Bacteria (BIB) in the remainder of the article. 

\subsection{Abundance estimation model in detail}

The statistical model for strain abundance estimation was based on a
statistical model of sequencing data as a mixture of reads from a set
of known reference sequences~\citep{Xing2006,Li2010}. The relative
proportions of the sequences are the unknown parameters $\theta$. In
our case the references were the core genomes of randomly selected
representatives of each cluster. Reads not mapping to the core genomes
were ignored.

After introducing indicator variables
$I_n$ defining the sequence of origin of each read $r_n$, the
likelihood of a read $r_n$ (single or paired-end) $p(r_n | I_n = m)$
is defined in Eq.~(1) of \citet{Glaus2012} and depends on the
mismatches in the alignment as well as the length of the reference
sequence.  The position model was not used in BIB because it would be
difficult to estimate with almost no unique alignments. We used a
conjugate $\mathrm{Dirichlet}(\alpha, \dots, \alpha)$ prior over
$\theta$ with $\alpha = 1$. Smaller $\alpha$ would mean weaker
regularisation, but $\alpha \ge 1$ is needed for log-concavity of the
model which aids convergence.

We used fast collapsed variational inference to optimise an approximate
posterior distribution over $I_n$ after marginalising out $\theta$
\citep{hensman2012fast,Hensman2015}. The posterior distribution over
the unknown proportions $\theta$ was obtained from these as in
\citet{Hensman2015}.

\subsection{Generation of data for validation experiments}

For the primary set of experiments, each sample was created by randomly mixing the reads from a number
of real single strain sequencing data sets described in
Table~\ref{tab:datasources} using fixed proportions.
These data are obtained independently of the reference sequences used
in the model and represent realistic sequencing data obtained from
other strains in the same clusters.  The data are available for
download on
Figshare\footnote{\url{http://dx.doi.org/10.6084/m9.figshare.1617539}}.

\begin{table}[htb]
  \caption{Sequencing data sets used to generate the mixed samples.}
  \label{tab:datasources}
  \centering
  \begin{tabular}{llll}
    Species & Strain & Accession & Cluster \\ \hline
    \emph{S. aureus} & ST5 & ERR107800 & 5 \\
    \emph{S. aureus} & ST8 & ERR107802 & 1 \\
    \emph{S. aureus} & ST30 & ERR107794 & 11 \\
    \emph{S. epidermidis} & TAW60 & dryad.85495~\citep{Meric2015} & 14 \\
    \emph{S. epidermidis} & CV28 & dryad.85494~\citep{Meric2015} & 15 \\
    \emph{S. epidermidis} & 1290N & dryad.85493~\citep{Meric2015} & 16 \\
    \emph{Bacillus subtilis} &  & DRR008449~\citep{Shiwa2013} & - \\
  \end{tabular}
\end{table}

To test more thoroughly the effect of dropped clusters in the presence of a more diverse representation of different strains, we additionally 
simulated reads using MetaSim~\citep{Richter2008}.

\section{Results}

We tested the BIB pipeline on several DNA sequencing data sets from
\emph{Staphylococcus} strains. We used two different types of data sets:
\begin{enumerate}
\item data sets with artificial mixtures of genuine reads from single strain
  sequencing experiments, and
\item synthetic data sets generated using MetaSim.
\end{enumerate}
We report the results
from our pipeline and compare against Pathoscope
2~\citep{Francis2013,Hong2014} as well as naive estimation from
strain frequencies among uniquely mapping reads. To ensure that the
other methods can fully utilise
the same information about the strains, we used the same read
alignments as input to all methods, essentially only replacing the
final abundance estimation step in our pipeline.

\subsection{Clustering and selection of strains}

\begin{figure}[htb]
  \centering
  \includegraphics[width=\textwidth]{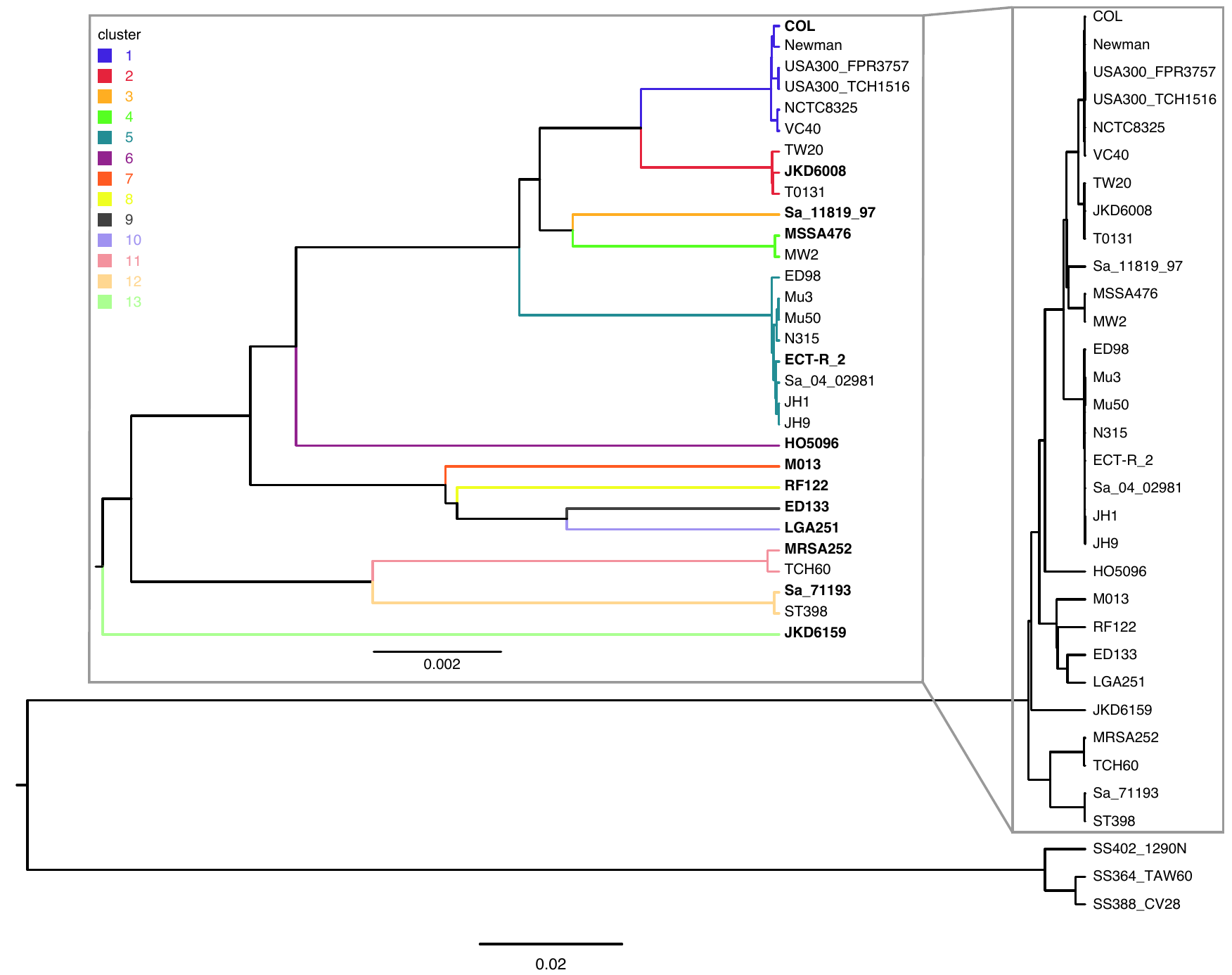}
  \caption{Phylogenetic tree of the investigated
    \emph{Staphylococcus} strains. Inset: Enlarged view of the \emph{S. aureus}
    branch illustrating the clustering of the strains within clonal complexes.
    The scale measures base-level sequence dissimilarity, showing that
    the \emph{S. aureus} clusters differ by approximately 2-10
    substitutions every 1 kb while strains within each cluster differ by
    less than 1 substitution every 5 kb.}
  \label{fig:tree}
\end{figure}

The strains used in the experiment and their phylogenetic
relationships are illustrated in Fig.~\ref{fig:tree}. The
phylogenetic tree illustrates the clonal complex (CC) structure of the
\emph{S. aureus} population~\citep{Feil2004}, where members
of the same complex are highly similar and interchangeable in terms of
strain identification~\citep{Meric2015}. Choosing one
representative for each CC corresponds to the clustering illustrated
in Fig.~\ref{fig:tree}.

\subsection{Identification of \emph{Staphylococcus} strains from sequencing data}

We generated 30 synthetic mixtures of sequencing
reads from different strains of \emph{Staphylococcus} species as
described and analysed these data sets
using BIB.  As a benchmark, we also tested the same identification and quantification using Pathoscope instead of
BitSeq.  Each analysed data set contained a mixture of 2-6
\emph{Staphylococcus} strains. The number of reads varied between
1--3 million.

Strain level identification is very
difficult, as typically only around $0.1--0.2 \%$ of the reads
map uniquely to the core genome.
Full genome alignments produce more unique hits, but
given the volatility of the accessory genome these are also likely to be
more misleading.

\begin{figure}[htb]
  \centering
  \includegraphics{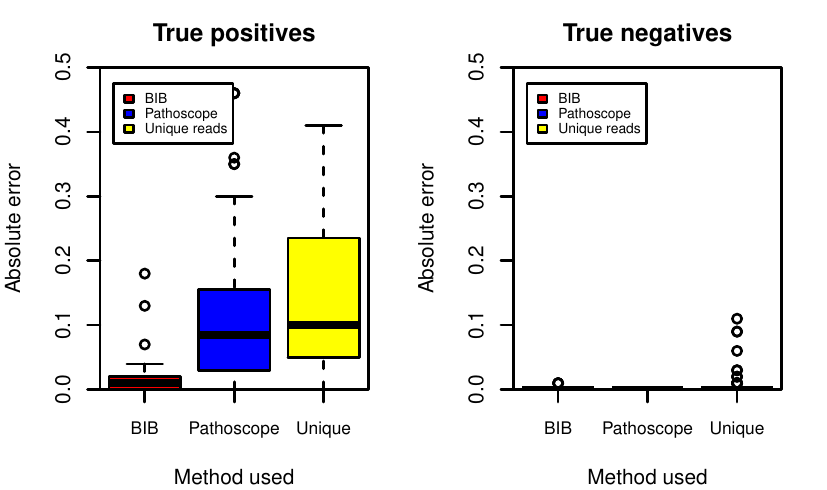}
  \caption{Magnitudes of errors in proportion estimates of BIB,
    Pathoscope and naive estimation among uniquely mapping reads
    (Unique) in strains really present in the experiment
    (true positives; left) and
    those not present in the experiment (true negatives; right).
    The ``Unique'' method is implemented by simply computing the
    frequencies of different strain clusters among unique alignments.
    Lower values indicate better results.}
  \label{fig:error_comparison}
\end{figure}

\begin{figure}[htb]
  \centering
  \includegraphics{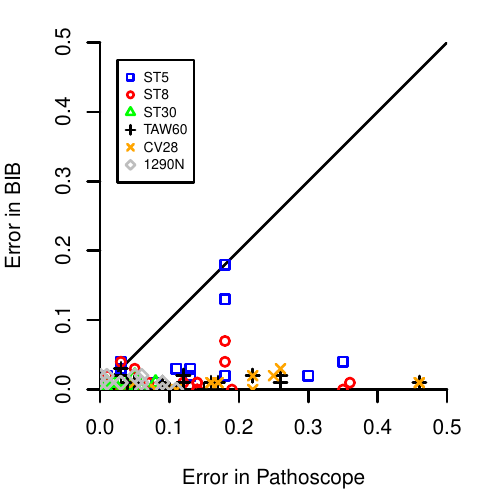}
  \caption{Scatter plot comparing the estimation errors of BIB and
    Pathoscope on true positives. Points below the diagonal are
    cases where BIB is more accurate while point above the diagonal
    are cases where Pathoscope is more accurate.}
  \label{fig:error_scatter}
\end{figure}

The absolute errors in the abundance estimation in the experiments are
illustrated in Fig.~\ref{fig:error_comparison}.  We split our analysis
to two cases: strains not present in the samples (true negatives) and
strains that are present (true positives).  All methods are
reliable in identifying true negatives.
For true positives, BIB
consistently provides very accurate quantification (absolute errors
mean $\pm$ standard deviation
$0.014 \pm 0.023$) while Pathoscope and the naive unique mapping read
analysis are significantly less accurate
(Pathoscope absolute errors $0.11 \pm 0.11$, unique reads
$0.14 \pm 0.12$).
BIB quantification results remain accurate all the way down to the
least abundant strains which had only 3 \% abundance in our data.
A scatter-plot in
Fig.~\ref{fig:error_scatter} comparing the errors of BitSeq and Pathoscope by each experiment
shows that BIB is essentially always more accurate than Pathoscope
($p < 10^{-16}$; Wilcoxon signed rank test) and often by a wide margin.

\subsection{Estimation in the presence contaminant species}

\begin{figure}[htb]
  \centering
  \includegraphics{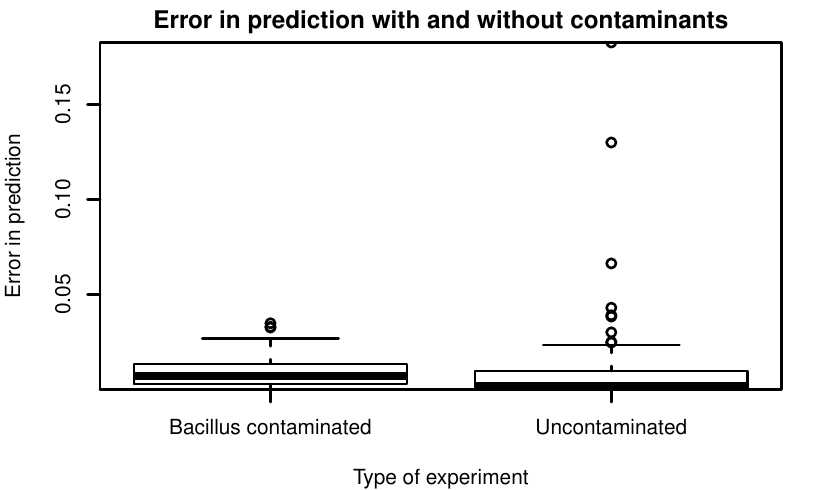}
  \caption{Comparison of errors in estimation of \emph{Staphylococcus}
    proportions with and without \textit{Bacillus} contamination.
    Errors on contaminated samples are slightly higher, but overall
    still very low.}
  \label{fig:contaminated}
\end{figure}

The alignment of reads to reference genomes makes BIB highly robust
to contamination by unrelated species. We tested this by
generating 10 of the samples with 3--30 \%
contamination from \textit{Bacillus subtilis subsp.\ subtilis} str.\ 168.
Full details of the experiments are given in Supplementary File 1.
After filtering the non-aligning reads, which include most of the \textit{Bacillus} reads, 
the estimation accuracy on \emph{Staphylococcus} proportions is
almost as good as with the uncontaminated samples, as illustrated in
Fig.~\ref{fig:contaminated}.  The corresponding median errors are
0.002 for the uncontaminated samples and 0.01 for the contaminated samples, respectively.
Addition of the contaminant reads is visible as a drop in the total
rate of aligned reads, but given the significant and variable number
of unmappable reads originating from the auxiliary genome the mapping
rate is at best an unreliable measure of the contamination level.

\subsection{Estimation in the presence of unknown strain clusters}

\begin{figure}[htb]
  \hspace{-25mm}\includegraphics{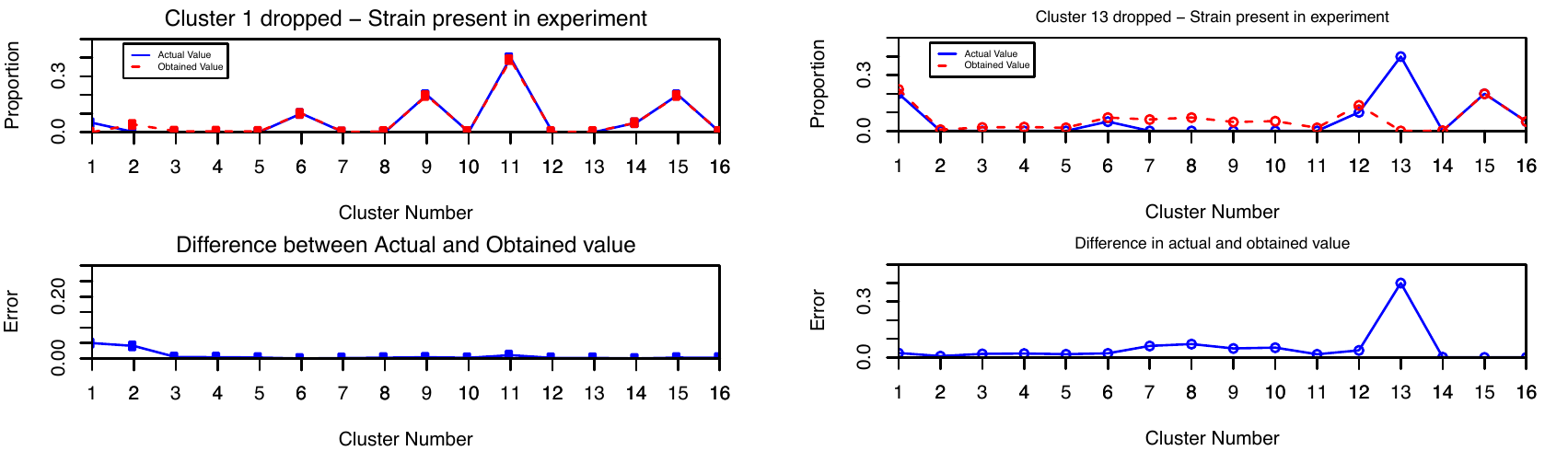}
  \caption{Two examples of error spectra when some strain
    clusters present in a sample are not included in the index. The plots show the profile
    of true and estimated proportions as well as the errors in the
    estimation. The lines will always show a bump at the dropped
    cluster index because they cannot be estimated.}
  \label{fig:dropped}
\end{figure}

When the reads of unknown origin stem from a species or strain related closely
enough to allow for the reads aligning well to those included in the index, they
tend to be assigned to the closest included reference strains. This is
illustrated by two examples in Fig.~\ref{fig:dropped}. In the first
example dropping Cluster 1 from the index causes the reads to get assigned to
Cluster 2 which is in the evolutionary sense closest to Cluster 1 in the phylogenetic tree in
Fig.~\ref{fig:tree}. In the second example, dropping Cluster 13, results in the
reads getting split more evenly among the available alternatives because the branch to Cluster 13 splits
off from the rest very early.

\subsection{Estimation without clustering}

\begin{figure}[htb]
  \centering
  \includegraphics{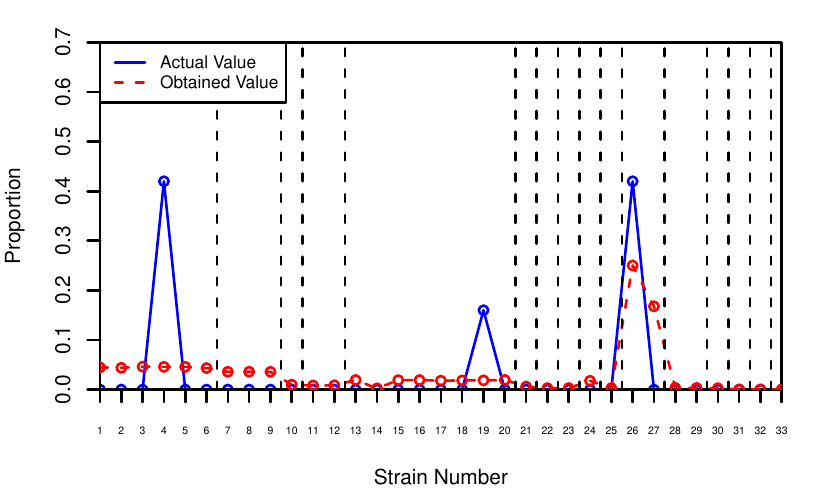}
  \caption{An example of error profile in strain abundance estimation
    without clustering. The vertical dotted lines indicate the borders
    between different clusters.}
  \label{fig:unclustered}
\end{figure}

Clustering of very similar strains when defining the reference set is an essential part of BIB.
Fig.~\ref{fig:unclustered} shows a typical example of the consequences of excluding the clustering step. As seen, the contribution
of a single cluster representative truly present in a sample tends to get
split up between all strains representing the same cluster in the reference set as they are too similar to
be differentiated. Furthermore, the method is unable to separate strains 1-9
belonging to Clusters 1 and 2, even though the two were usually
properly separated in the experiments with clustering of the reference strains. This is most likely
because the difference of using 6 or 9 strains to represent the data
is not as substantial as the difference between 1 or 2 strain clusters where
the clearly simpler model is able to drive the other coefficient to
zero. It is likely that no statistical method would be able to truthfully resolve origins of reads when the sources are too similar to each other. Hence, it is of importance to ensure biological meaningfulness of the reference set of strains prior to the assignment analysis. 

\subsection{Analysis of clinical samples}

\begin{figure}[htb]
  \centering
  \includegraphics{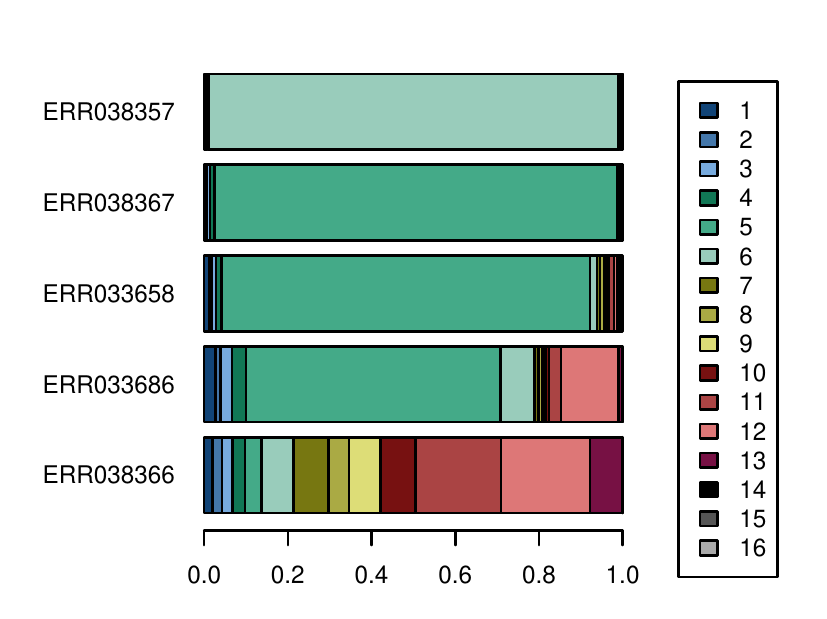}
  \caption{Estimated cluster abundance profiles from diverse clinical
    samples. The two top rows represent clean samples where one
    cluster clearly dominates. Rows 3 and 4 represent contaminated
    samples where the true cluster can still be fairly reliably
    identified. The bottom row shows a completely failed sample,
    possibly due to problems with sequence barcoding.}
  \label{fig:clinical}
\end{figure}

To illustrate the practical applicability of BIB we tested it on
\emph{S.\ aureus} short read data generated at the Wellcome Trust
Sanger Institute as part of a Europe-wide surveillance project
(``Genetic diversity in \emph{Staphylococcus aureus} (European
collection)'' study), with kind permission from Matthew
Holden. Initial analysis of these data revealed they were of poor
quality, probably resulting from contamination, and for this reason
they have not previously been published. All isolates were recovered
from cases of invasive \emph{S. aureus} disease. The estimated
abundance profiles of selected samples are shown in
Fig.~\ref{fig:clinical}. In top two isolates (ERR038357 and ERR038367)
a single cluster is robustly identified ($>95\%$ share for the
dominant strain, all other shares $<1\%$) indicating that the level of
contamination in these samples is low. In contrast, isolates ERR033658
and ERR033686 (rows 3 and 4) show clearer evidence of mixed clusters
due to contamination. We also note that the cluster profiles are
similar within these two samples, which is consistent with a single
source of contamination for both runs. Isolate ERR038366 (bottom row)
represents a completely failed sample, possibly caused by problems
with sequence barcoding.

\subsection{Runtime}

For a new sample, the pipeline requires running programs for read
alignment (Bowtie 2) and abundance estimation (BitSeq being the core part of BIB). The time required
by these two steps is approximately equal. A typical analysis of 1 M
reads takes approximately 10 min on a single CPU desktop computer
representing a standard level of hardware.

\section{Discussion}

\subsection{Interpretation of proportions and benchmarking}

Our pipeline can estimate strain abundances as proportions of the
sequencing reads. These would be expected to be related to the
proportions of DNA from the different strains. Depending on the
relative lengths of different genomes, this may deviate slightly from
cell counts between species, but should be consistent within a species
because we only consider the shared core genome of equal length. This
kind of minor variations should not affect any practical applications.

Our empirical evaluation is based solely on synthetic mixtures of
sequencing reads from different single strain sequencing experiments.
Such mixtures are necessary to enable accurate benchmarking of
the methods.  Because we use actual reads from various experiments
they will not perfectly match the reference and thus represent a much
more realistic test than synthetic reads generated from references.
Experiments based on laboratory derived mixed cultures would add
significant extra uncertainty because it is difficult to accurately control
the strain proportions during the cultivation process.

\subsection{Applicability to different bacterial species}

The main assumption behind our BIB method is that each putative biologically meaningful source is adequately represented by a single core genome sequence to which the reads can be mapped.
As illustrated in this paper, this works with high fidelity for species like
\emph{S. aureus} whose population structure has clear well-separated
lineages~\citep{Meric2015}. Preliminary experiments suggest the method
may not work as well for species experiencing more frequent recombination. Extension
of the work to such species is an important avenue of future work.
The current state-of-the-art probabilistic identification method Pathoscope
2~\citep{Francis2013,Hong2014} is essentially based on a similar assumption and is expected to be similarly vulnerable to strong deviations from the assumption. However, our experiments demonstrated that BIB delivers a considerably higher level of estimation accuracy without requiring more extensive computational resources.

As illustrated in the results of Fig.~\ref{fig:unclustered},
clustering the strains is essential for accurate identification
results. It is not surprising that
distinguishing among multiple highly related strains is infeasible,
however, it is more striking that clustering also aids identification of read origin
between the more separated sources. We suspect this may be due to the prior used in the
Bayesian model, but further work is needed to properly understand the
phenomenon.

In transcript-level RNA-seq analysis clustering of similar transcripts
has been suggested for improving the accuracy by~\citet{Turro2014}.
Unlike our off-line clustering, their algorithm is run on-line
together with the inference separately for every sample.
Our approach can easily incorporate additional expert knowledge and
guarantee consistent clustering making interpretation of the results
more straightforward.
This approach is expected to work especially well for any species that has a clear subpopulation boundaries, since every
potentially mixed sample will correspondingly have a clearly delineable structure among its reads, apart from those representing contamination which can be efficiently filtered out by our pipeline.

\subsection{Relationship to transcript-level RNA-seq analysis}

The underlying statistical problem in bacterial strain identification
is the same as that underlying most transcript-level RNA-seq
expression estimation methods: how to estimate the probability of a
read originating from a given reference sequence.  There exist a
number of methods solving the same problem there including
RSEM~\citep{Li2010,Li2011}, Cufflinks~\citep{Trapnell2010},
Miso~\citep{Katz2010}, BitSeq~\citep{Glaus2012,Hensman2015},
TIGAR~\citep{nariai2013tigar,Nariai2014},
eXpress~\citep{roberts2013streaming}, Sailfish~\citep{patro2014sailfish}
and many others. These are all
based on different inference methods applied to the same probabilistic
model first proposed in~\citep{Xing2006}.  This is also essentially the
same as the model used by Pathoscope~\citep{Francis2013,Hong2014}.
There are also a number of other RNA-seq analysis methods based on
other models.  We have chosen to use the fast variational Bayes (VB) version of
BitSeq~\citep{Hensman2015} as core ingredient in BIB because it provides very high accuracy while
being reasonably fast according to recent broad
assessments~\citep{SEQC2014,Kanitz2015}.

\section{Conclusion}

In this paper we have presented the BIB pipeline for probabilistic identification and
quantification of relative abundance of bacterial strains in mixed samples from unassembled sequence data.  The pipeline is based on
alignment of reads to representative core genomes followed by deconvolution of multi-mapping reads using BitSeq,
a method previously proposed for RNA-seq analysis.  Our BIB pipeline 
can rapidly and reliably estimate the proportions of the reference strains with the typical
deviance of at most a few percent units, using approximately 1 M sequencing reads. BIB improves
significantly upon the accuracy of both naive analysis as
well as previous state-of-the-art method in strain identification.
Application of BIB to analyse clinical samples suggests it has
significant potential both in strain identification as well as
flagging problematic, such as contaminated, samples.

\subsubsection*{Acknowledgements}

Some of the data used in this study were provided by the Microorganism
group at the Wellcome Trust Sanger Institute and can be obtained from
the European Nucleotide Archive.

\subsubsection*{Funding}

This work was supported by the Academy of Finland [259440 to A.H.,
251170 to J.C.]; the European Research Council [239784 to J.C.]; and
the Medical Research Council [MR/L015080/1 to S.K.S.].

\bibliography{../NAR2015/bactstrainid_refs}
\bibliographystyle{myabbrvnat}

\end{document}